\documentclass[final,1p,times,twocolumn]{elsarticle}

\usepackage{amssymb}
\usepackage{amsmath}
\usepackage{amsthm}

\journal{Physical Communication}

\begin{document}

\begin{frontmatter}



\title{Joint Physical Network Coding and LDPC decoding for Two Way Wireless Relaying}
\tnotetext[t1]{This work was supported by the National Natural
Science Foundation of China (No. 60972050) and the Major Special
Project of China under Grant (2010ZX03003-003-01) and the Jiangsu
Province National Science Foundation under Grant
(BK2011002,BK2012055).}

\author{Kui Xu, Zhenxing Lv, Youyun Xu, Dongmei Zhang, Xinyi Zhong, Wenwen Liang}

\address{Institute of Communications Engineering, PLAUST, Nanjing, 210007}

\begin{abstract}
In this paper, we investigate the joint design of channel and
network coding in bi-directional relaying systems and propose a
combined low complexity physical network coding and LDPC decoding
scheme. For the same LDPC codes employed at both source nodes, we
show that the relay can decodes the network coded codewords from the
superimposed signal received from the BPSK-modulated multiple-access
channel. Simulation results shown that this novel joint physical
network coding and LDPC decoding method outperforms the existing MMSE
network coding and LDPC decoding method over AWGN and complex MAC channel.

\end{abstract}

\begin{keyword}
Joing Network Coding and LDPC Decoding; Two Way Wireless Relaying
Channel; Wireless Cooperative Networks.

\end{keyword}

\end{frontmatter}


\section{Introduction}
\label{} The network coding scheme was originally considered as a
technique of improving network throughput for wired networks [1]. In
wireless network, the broadcast nature of the wireless channel is
usually considered to cause enormous interference if more than two
nodes transmit simultaneously at the same frequency. On the other
hand, physical network coding (PNC) [2,3] can employ this broadcast
nature as a capacity-boosting approach for two-way or multi-way
cooperative communication network.

A simple two-way wireless relaying system with two sources A and B and one relay R is depicted in Fig.1. Source A and source B desire to exchange information between each other and there is no direct link between the two source nodes. Thus, all the transmission between source A and B must flow through the relay R. The relay transmission consists of two states: multiple access (MAC) stage, where source A and B transmit the LDPC-coded signals to the relay R simultaneously, and broadcast (BC) stage, where the relay R broadcasts to both source A and B. One critical process at R is to decode the superimposed signal from A and B at MAC stage in such a way that A and B could decode the information from each other reliably at the BC stage. Instead of decoding the individual information belonging to the source A and B separately, the relay node R aims to decode the received superimposed signal to the network-coded combination of the two sources information. We refer this decoding process as the joint physical network coding and LDPC decoding (JNCLD).

\begin{figure}[h]
\begin{center}
\includegraphics[scale=0.7]{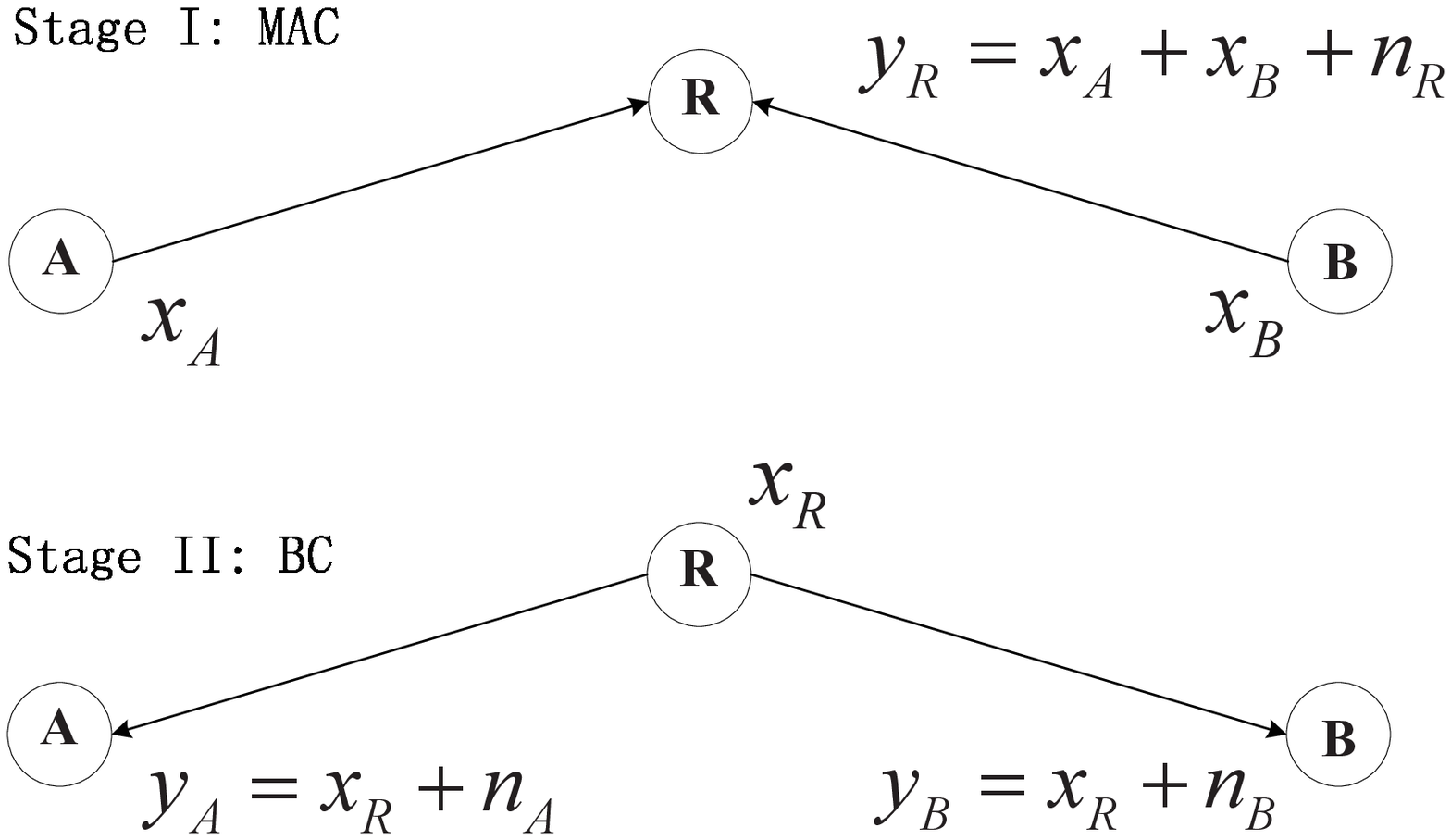}
\end{center}
\caption{ Two way relay channel}
\label{fig 1}
\end{figure}

In [4,5] non-coherent physical layer network coding for FSK and
CPFSK modulation are proposed. In [6], joint network and channel
coding was proposed for the simple real additive multiple-access
white Gaussian noise channel. By noticing the linearity of both
network and channel coding, the soft Log-likelihood Ratios (LLRs)
for the network-coded codeword can be directly estimated from the
received physically superimposed signals. In [7,8], a joint network
and LDPC coding scheme for bi-directional relaying is presented and
the closed-form expressions for computing the log-likelihood ratios
of the network-coded codewords have been derived for both real and
complex multiple access channels. An adaptive PNC called pseudo
exclusive-or (PXOR) for LDPC coded two-way relay block fading
channels is proposed in [12]. Based on the pairwise check decoding
(PCD) [13], the check relationship table generated by PXOR mapping
obtains the same Hamming distances as that of conventional XOR
mapping. In order to compensate the amplitude fading and phase
deviation of the TWR block fading channels, the PXOR mapping
optimizes the Euclidean distances by adjusting the symbol distances
dynamically.

Optimal time and rate allocation scheme for a network-coded
bidirectional wireless communication system is proposed in [14]. The
closed form expressions for the optimal allocation of the
transmission time and of the data rate in both ways for given
channel SNRs to maximizing the sum-rate is derived. In [9,10], a
joint PNC and minimum mean square error (MMSE) based LDPC decoding
method is proposed. In this paper, we provide some novel insights
into the decode-and-forward approach for two way wireless relaying
with BPSK signaling. In particular, for the same LDPC codes employed
at both source nodes, a novel iterative LDPC decoding algorithm is
proposed for the physical network coding scheme at relay, which
outperforms recently proposed MMSE based network coding scheme on
bit error rate (BER) performance.

The remainder of this paper is organized as follows. In Section II
we present the considered system model. Section III present the
joint PNC and MMSE based LDPC decoding scheme [9,10] and introduce
the proposed novel joint physical network coding and LDPC decoding
approach. Finally, in Section IV and V simulation results are
presented and conclusion is given.

\section{System Model}
We denote $b_{A}\in\{0,1\}^{K} $ and $b_{B}\in\{0,1\}^{K} $ as the information vector of the two source nodes A and B, respectively. The information is encoded by the same LDPC code with a code rate of $K/N$ into the codeword vectors $c_{A}\in\{0,1\}^{N} $ and $c_{B}\in\{0,1\}^{N} $ at the sources A and B, respectively. The encoded vectors are BPSK-modulated to $x_{A}\in\{-1,1\}^{N} $ and $x_{B}\in\{-1,1\}^{N} $  according to the mapping rule $1\rightarrow1$ and $0\rightarrow-1$.

\subsection{Multiple Access Stage}
During the MAC stage, the two sources transmit the modulated signals $x_{A}$ and $x_{B}$ to the relay R simultaneously. Under a multiple access white Gaussian noise channel and the assumption of perfect synchronization, the received signal at the relay R is
\begin{equation}
y_{R}=x_{A}+x_{B}+n_{R}
\end{equation}
where $n_{R}$  are identically distributed (i.i.d) zero-mean
Gaussian random variables with variance $\sigma^{2}_{n}$. According
to the physical network coding scheme introduced in [6-11], the XOR
of the source information denoted by $b_{A\oplus B}=b_{A}\oplus
b_{B}\in\{0,1\}^{K}$ can be estimated at the relay from the received
signal $y_{R}$, i.e., $b_{R}=\hat{b}_{A \oplus B}\in\{0,1\}^{K}$.
Then, $b_{R}$ is encoded by the same LDPC code, and the code vector
$c_{R}$ BPSK-modulated to $x_{R}$.
\subsection{Broadcasting Stage}
In the BC stage, the relay R broadcasts $x_{R}$ to both A and B. Thus, the received signals at A and B are given by
\begin{equation}
\begin{array}{*{20}{c}}
y_{A}=x_{R}+n_{A}\\
y_{B}=x_{R}+n_{B}
\end{array}
\end{equation}

At both A and B, the information $\hat{b}_{R}$ , which contains the information of $b_{A \oplus B}$, is estimated from $y_{A}$ and $y_{B}$, respectively. Since A and B know what has been transmitted at the MAC stage, A and B can obtain the information from each other simply by means of the binary XOR, i.e., $\hat{b}_{B}=\hat{b}_{R}\oplus\hat{b}_{A}$ and $\hat{b}_{A}=\hat{b}_{R} \oplus \hat{b}_{B}$.

A critical process at the relay R is to decode the superimposed signal from A and B in such a way that A and B could decode the information from each other reliably at the BC stage. In this paper, we will focus on deriving a decoding algorithm for $y_{R}\rightarrow b_{R}$.

\section{Joint Physical Network Coding and LDPC Decoding}
In [9,10], joint PNC and MMSE based LDPC decoding method is
proposed. The idea of this algorithm is that under the assumption of
the same LDPC code applied at both source nodes, i.e., parity check
matrix $H_{A}=H_{B}=H$ and
\begin{equation}
\begin{array}{*{20}{c}}
H_{A}c_{A}=0,\\
H_{B}c_{B}=0,\\
H(c_{A} \oplus c_{B})=0
\end{array}
\end{equation}
the XOR of the encoded symbols $c_{A \oplus B} \in \{0,1\}^{N}$ is also a valid codeword of the LDPC code. The relay R first maps each pair of the received superimposed signal $y_{R}$ to an estimation of the joint network and LDPC coded symbol $\hat{y}_{R}^{MMSE}$  corresponding to $c_{A} \oplus c_{B}$ by using minimum mean square error (MMSE) estimation; then performs LDPC decoding on the interim symbol $c_{A} \oplus c_{B}$ to obtain the network coded symbol $b_{A} \oplus b_{B}$.

Different from the above mentioned approach, we will compute the LLR
$\Lambda(c_{A \oplus B})$ of the XOR of the two source information
$c_{A} \oplus c_{B}$ direct from the received signal $y_{R}$ and
then performs LDPC decoding to obtain the network coded symbol
$b_{A} \oplus b_{B}$.

\subsection{AWGN MAC Channel with Power Allocation}
The received superimposed signal $y_{R}$ over AWGN MAC channel with power allocation can be expressed as
\begin{equation}
y_{R}=h_{A}\rho_{A}x_{A}+h_{B}\rho_{B}x_{B}+n_{R}
\end{equation}
where $h_{A} $ and $h_{B} $ represent the channel gain from node A
and B to the relay node R, respectively. $P_{A}=\rho_{A}^{2}$ and
$P_{B}=\rho_{B}^{2}$ denote the power allocated to the MAC channel,
respectively. Let $n$ index the bit of a codeword. The a-priori
probabilities of $\{c_{A \oplus B} (n)=c_{A}(n) \oplus c_{B}(n)=0,1
\}$ are
\begin{equation}
\begin{array}{*{20}{c}}
Pr\{c_{A \oplus B}(n)=0\}=1/2\\
Pr\{c_{A \oplus B}(n)=1\}=1/2\\
\end{array}
\end{equation}

If $c_{A \oplus B}(n)=1$ holds, the event $E_{1}=\{c_{A}(n)=0,c_{B}(n)=1\}$ or the event $E_{2}=\{c_{A}(n)=1,c_{B}(n)=0\}$ should be satisfied. On the other side, if $c_{A \oplus B}(n)=0$ holds, the event $E_{3}=\{c_{A}(n)=1,c_{B}(n)=1\}$ or the event $E_{4}=\{c_{A}(n)=0,c_{B}(n)=0\}$ should be satisfied. Hence, the probability of $\{c_{A \oplus B}(n)=1\}$  is the sum probability of $Pr\{E_{1}\}$ and $Pr\{E_{2}\}$ , and the probability of $\{c_{A \oplus B}(n)=0\}$  is the sum probability of $Pr\{E_{3}\}$ and $Pr\{E_{4}\}$.

Assuming that the relay has knowledge of parameter $\sigma_{R}^{2}$ and channel state information. Soft-decision decoding requires that the relay node R compute the LLR of each network coded bit $c_{A \oplus B}$ according to
\begin{eqnarray}
\Lambda_{AM}(c_{A \oplus B}) &=& log\frac{P(c_{A \oplus B}=1|y_{R})}{P(c_{A \oplus B}=0|y_{R})} \nonumber\\
&=& log\frac{P(c_{A} \oplus c_{B}=1|y_{R})}{P(c_{A} \oplus c_{B}=0|y_{R})}\\
&=& log[P(y_{R}|E_{1})+P(y_{R}|E_{2})]-log[P(y_{R}|E_{3})+P(y_{R}|E_{4})] \nonumber
\end{eqnarray}

where
\begin{equation}
P(y_{R}|E_{1})=\frac{1}{4\sqrt{2 \pi
\sigma_{R}^{2}}}exp\Bigg\{-\frac{(y_{R}-h_{A}\rho_{A}+h_{B}\rho_{B})^{2}}{2\sigma_{R}^{2}}\Bigg\}
\end{equation}
\begin{equation}
P(y_{R}|E_{2})=\frac{1}{4\sqrt{2 \pi
\sigma_{R}^{2}}}exp\Bigg\{-\frac{(y_{R}+h_{A}\rho_{A}-h_{B}\rho_{B})^{2}}{2\sigma_{R}^{2}}\Bigg\}
\end{equation}
\begin{equation}
P(y_{R}|E_{3})=\frac{1}{4\sqrt{2 \pi
\sigma_{R}^{2}}}exp\Bigg\{-\frac{(y_{R}-h_{A}\rho_{A}-h_{B}\rho_{B})^{2}}{2\sigma_{R}^{2}}\Bigg\}
\end{equation}
\begin{equation}
P(y_{R}|E_{4})=\frac{1}{4\sqrt{2 \pi
\sigma_{R}^{2}}}exp\Bigg\{-\frac{(y_{R}+h_{A}\rho_{A}+h_{B}\rho_{B})^{2}}{2\sigma_{R}^{2}}\Bigg\}
\end{equation}

The LLR of the network coded bit $c_{A \oplus B}$ over AWGN MAC channel with power allocation can be expressed as
\begin{eqnarray}
\Lambda_{AM}(c_{A \oplus B}) &=& \frac{2h_{A}h_{B}\rho_{A}\rho_{B}}{\sigma_{R}^{2}}+log\Bigg(cosh\bigg(\frac{y_{R}(h_{A}\rho_{A}-h_{B}\rho_{B})}{\sigma_{R}^{2}}\bigg)\Bigg)\\
&-&log\Bigg(cosh\bigg(\frac{y_{R}(h_{A}\rho_{A}+h_{B}\rho_{B})}{\sigma_{R}^{2}}\bigg)\Bigg)\nonumber
\end{eqnarray}

The network coded information bit $b_{A \oplus B}$ can be obtained by the traditional belief broadcasting (BP) decoding algorithm with LLR
$\Lambda_{AM}(c_{A \oplus B})$.

\subsection{Complex MAC channel with Power Allocation}
The received superimposed signal $\tilde{y}_{R}$ over complex MAC channel with power allocation can be expressed as
\begin{equation}
\tilde{y}_{R}=h_{A}e^{j\theta_{A}}\rho_{A}x_{A}+h_{B}e^{j\theta_{B}}\rho_{B}x_{B}+n_{R}
\end{equation}
where $n_{R}$ denotes the zero-mean complex Gaussian noise for
complex MAC channel with covariance $\sigma_{R}^{2}$, $h_{A}$ and
$h_{B}$ represent the channel gain from node A and B to the relay
node R, respectively. $P_{A}=\rho_{A}^{2}$ and $P_{B}=\rho_{B}^{2}$
denote the power allocated to the complex MAC channel, respectively.
$\theta_{A}$ and $\theta_{B}$ represents uniformly distributed phase
shift over $[0,2\pi)$. The LLR of each network coded bit $c_{A
\oplus B}$ with the assumption that the relay has knowledge of
parameter $\sigma_{R}^{2}$ and channel state information can be
expressed as
\begin{eqnarray}
\Lambda_{CM}(c_{A \oplus B}) &=& log\frac{P(c_{A \oplus B}=1|\tilde{y}_{R})}{P(c_{A \oplus B}=0|\tilde{y}_{R})} \nonumber\\
&=& log\frac{P(c_{A} \oplus c_{B}=1|\tilde{y}_{R})}{P(c_{A} \oplus c_{B}=0|\tilde{y}_{R})}\\
&=& log[P(\tilde{y}_{R}|E_{1})+P(\tilde{y}_{R}|E_{2})]-log[P(\tilde{y}_{R}|E_{3})+P(\tilde{y}_{R}|E_{4})] \nonumber
\end{eqnarray}

where
\begin{equation}
P(\tilde{y}_{R}|E_{1})=\frac{1}{4\sqrt{2 \pi
\sigma_{R}^{2}}}exp\Bigg\{-\frac{\|\tilde{y}_{R}-h_{A}e^{j\theta_{A}}\rho_{A}+h_{B}e^{j\theta_{B}}\rho_{B}\|^{2}}{2\sigma_{R}^{2}}\Bigg\}
\end{equation}
\begin{equation}
P(\tilde{y}_{R}|E_{2})=\frac{1}{4\sqrt{2 \pi
\sigma_{R}^{2}}}exp\Bigg\{-\frac{\|\tilde{y}_{R}+h_{A}e^{j\theta_{A}}\rho_{A}-h_{B}e^{j\theta_{B}}\rho_{B}\|^{2}}{2\sigma_{R}^{2}}\Bigg\}
\end{equation}
\begin{equation}
P(\tilde{y}_{R}|E_{3})=\frac{1}{4\sqrt{2 \pi
\sigma_{R}^{2}}}exp\Bigg\{-\frac{\|\tilde{y}_{R}-h_{A}e^{j\theta_{A}}\rho_{A}-h_{B}e^{j\theta_{B}}\rho_{B}\|^{2}}{2\sigma_{R}^{2}}\Bigg\}
\end{equation}
\begin{equation}
P(\tilde{y}_{R}|E_{4})=\frac{1}{4\sqrt{2 \pi
\sigma_{R}^{2}}}exp\Bigg\{-\frac{\|\tilde{y}_{R}+h_{A}e^{j\theta_{A}}\rho_{A}+h_{B}e^{j\theta_{B}}\rho_{B}\|^{2}}{2\sigma_{R}^{2}}\Bigg\}
\end{equation}

The LLR of the network coded bit $c_{A \oplus B}$ over complex MAC channel with power allocation can be expressed as

\begin{eqnarray}
\Lambda_{AM}(c_{A \oplus B}) &=& \frac{\|h_{A}e^{j\theta_{A}}\rho_{A}+h_{B}e^{j\theta_{B}}\rho_{B}\|^{2}-\|h_{A}e^{j\theta_{A}}\rho_{A}-h_{B}e^{j\theta_{B}}\rho_{B}\|^{2}}{\sigma_{R}^{2}}\nonumber\\
&+&log\Bigg(cosh\bigg(\frac{\Re\big(\tilde{y}_{R}(h_{A}e^{j\theta_{A}}\rho_{A}-h_{B}e^{j\theta_{B}}\rho_{B})^{\ast}\big)}{\sigma_{R}^{2}}\bigg)\Bigg)\\
&-&log\Bigg(cosh\bigg(\frac{\Re\big(\tilde{y}_{R}(h_{A}e^{j\theta_{A}}\rho_{A}+h_{B}e^{j\theta_{B}}\rho_{B})^{\ast}\big)}{\sigma_{R}^{2}}\bigg)\Bigg)\nonumber
\end{eqnarray}

The network coded information bit $b_{A \oplus B}$ can be obtained by the traditional BP decoding algorithm with LLR
$\Lambda_{CM}(c_{A \oplus B})$.

\section{Simulation Results}
In this section, we demonstrate the simulated performance of the
proposed joint physical network coding and LDPC decoding scheme for
two way wireless relay systems with power allocation in each source
node. In the simulation, we assume that the SNR at the relay node R
is defined as $(P_{A}h_{A}^2+P_{B}h_{B}^2)/\sigma_{R}^{2}$ for both
AWGN and complex MAC channel, where $\sigma_{R}^{2}$ denotes the
noise variance received at the relay R and the total transmitting
power $P_{A}h_{A}^2+P_{B}h_{B}^2$ is set to 2. We check the BER (bit
error rate) of the decoded packet $b_{A \oplus B}=b_{A} \oplus
b_{B}$ at the relay node R. BPSK modulation scheme is used at both
end nodes for all simulations.

For comparison, we also study the performance of MMSE based method
proposed in [9,10]. We first compare the BER performance of MMSE
estimation and the proposed method over AWGN MAC channel with power
allocation for various packet length. In Fig. 2, the ration of
$P_{A}h_{A}^2/P_{B}h_{B}^2$ is set to be 2/3 and the iteration
numbers of the two schemes are set to 30. The proposed method
outperforms MMSE about 0.2dB when the BER is $10^{-4}$ for the
packet length (PL) of 1010 over AWGN MAC channel. The BER
performance of MMSE estimation and the proposed method over complex
MAC channel is given in Fig. 3, the power ration of the two source
nodes $P_{A}h_{A}^2/P_{B}h_{B}^2$ is also set to be 2/3. We can see
from Fig.3 that the proposed method outperforms MMSE based method
about 0.1dB when the BER is $10^{-4}$ for the PL of 1010.

\begin{figure}[h]
\begin{center}
\includegraphics[scale=0.7]{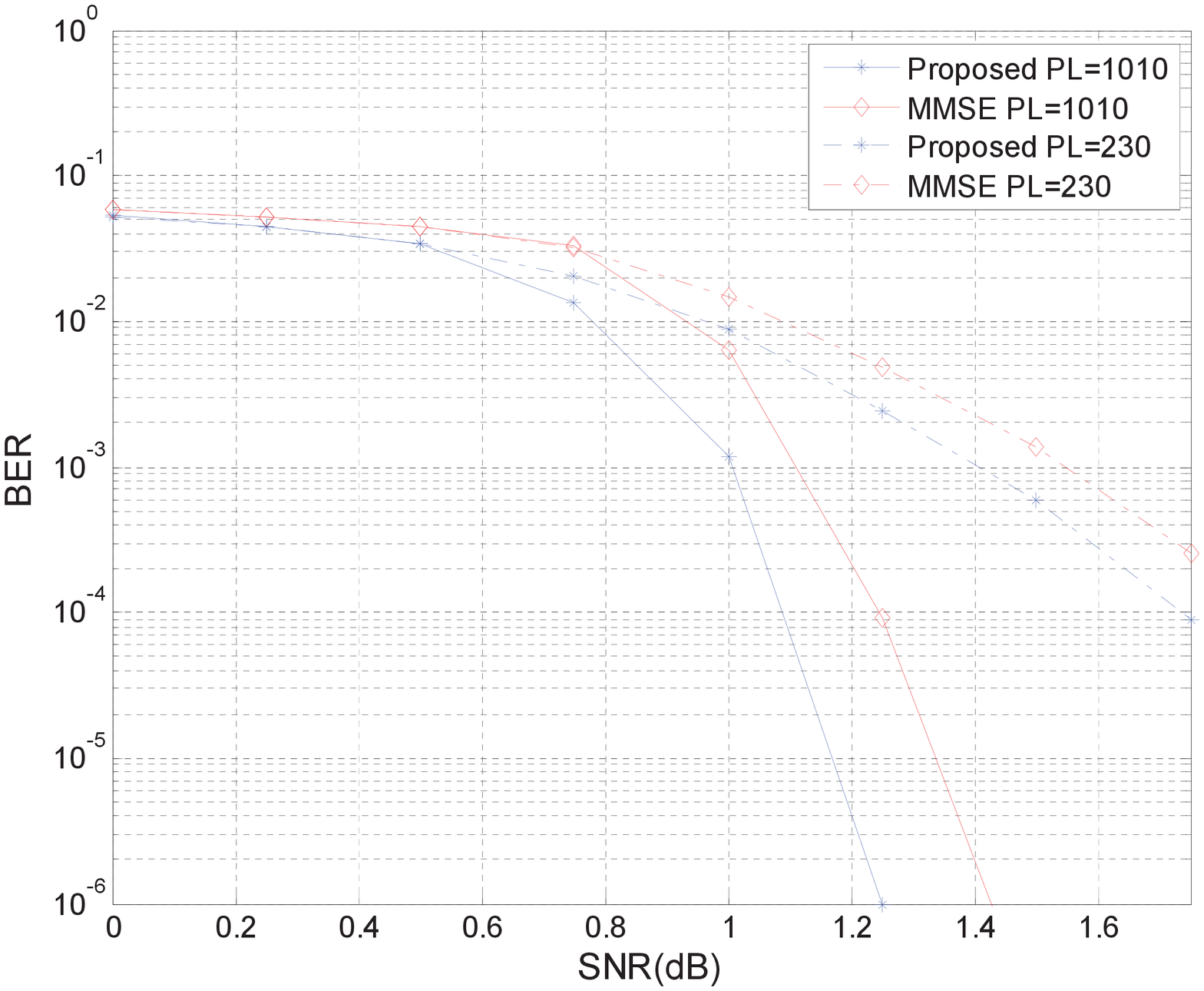}
\end{center}
\caption{BER performance of the proposed method and MMSE method for various packet lengths and power allocation over AWGN MAC channel}
\label{fig 2}
\end{figure}

\begin{figure}[h]
\begin{center}
\includegraphics[scale=0.7]{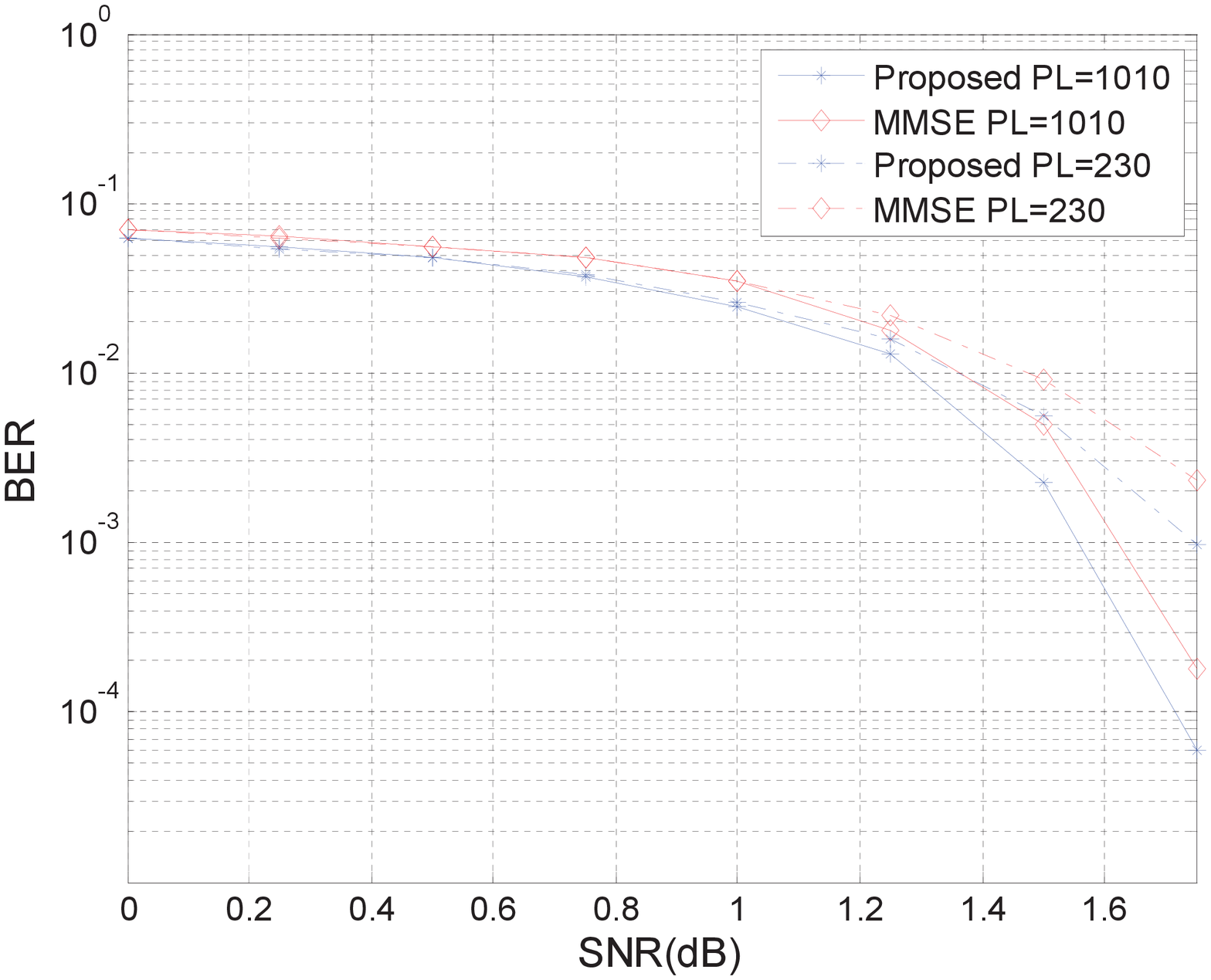}
\end{center}
\caption{BER performance of the proposed method and MMSE method for various packet lengths with power allocation and uniformly distributed phase shift caused by complex MAC channel}
\label{fig 3}
\end{figure}

\section{Conclusion}
In this paper, a novel joint physical network coding and LDPC decoding method for two way wireless relaying system with BPSK signaling is presented. The proposed method employ the same LDPC codes at the source nodes and the relay decodes the network coded packet from received superimposed signal by using the proposed method. Simulation results shown that the proposed novel method outperforms MMSE method about 0.2dB and 0.1dB over AWGN and complex MAC channel, respectively.


\begin{thebibliography}{99}


\bibitem{1}
R. Ahlswede, N. Cai, S.-Y. R. Li, and R. W. Yeung, Network
Information Flow, \emph{IEEE Trans. on Information Theory}, 46(4), pp. 1204-1216, July 2000.
\bibitem{2}
C. Hausl and J. Hagenauer, Iterative Network and Channel Decoding
for the Two-way Relay Channel, in \emph{Proc. IEEE International
Conference on Communications (ICC)}, Istanbul, Turkey, June 2006.
\bibitem{3}
P. Popovski and H. Yomo, Physical Network Coding in Two-Way Wireless
Relay Channels, in \emph{Proc. IEEE International Conference on
Communications (ICC)}, Glasgow, Scotland, June 2007.
\bibitem{4}
J. H. S{\o}rensen, R. Krigslund, P. Popovski, et al, Physical Layer Network Coding for FSK Systems, \emph{IEEE Communications Letters}, 2009, 13, (8), pp. 597-599
\bibitem{5}
M.C. Valenti, D. Torrieri and T. Ferrett, Noncoherent Physical-Layer
Network Coding, in \emph{Proc. IEEE Military Communications
Conference}, 2009.
\bibitem{6}
A. Zhan and C. He, Joint design of channel coding and physical
network coding for wireless network, in \emph{Proc. IEEE
International Conference on Neural Networks and Signal Processing
(ICNNSP)}, Zhe Jiang, China, Jun. 2008.
\bibitem{7}
Xiaofu Wu, Weijun Zeng, Chunming Zhao, and Xiaohu You, Joint Network
and LDPC Coding for Bi-directional Relaying. in \emph{Proc. IEEE
International Conference on Information Theory and Information
Security (ICITIS)}, Hang Zhou, China, Nov. 2011.
\bibitem{8}
Kui Xu, Youyun Xu, Wenwen Liang, Dongmei Zhang, Zhenxing Lv, Joint
LDPC and physical network coding with power allocation for two way
wireless relaying, in \emph{Proc. IEEE International Conference on
Wireless Communications and Signal Processing (WCSP)}, Nanjing,
China, Nov. 2011.
\bibitem{9}
S. Zhang, S. Liew, Channel Coding and Decoding in a Relay System
Operated with Physical-Layer Networking Coding. in \emph{Proc. IEEE
Journal on Selected Areas in Communications}, 2009, 27, (5), pp.
788-796.
\bibitem{10}
S. Zhang, S. Liew and L. Lu, Physical Layer Networking Coding
Schemes over Finite and Infinite Fields, in \emph{Proc. IEEE
GLOBECOM}, 2008.
\bibitem{11}
S. Zhang, S. Liew, and P. Lam, Physical Layer Network Coding, in
\emph{Proc. International Conference on Mobile Computing and
Networking (MobiCom)}, Los Angeles, USA, 2006, pp. 358-365.
\bibitem{12}
J. Liu, M. Tao, and Y. Xu, Pseudo Exclusive-OR for LDPC Coded
Two-Way Relay Block Fading Channels, in \emph{Proc. IEEE
International Conference on Communications (ICC)}, Kyoto, Japan,
June 2011.
\bibitem{13}
J. Liu, M. Tao, Y. Xu, and X. Wang, Pairwise check decoding for LDPC
coded two-way relay fading channels, in \emph{Proc. IEEE
International Conference on Communications (ICC)}, Cape Town, South
Africa, May 2010.
\bibitem{14}
C. Hausl, O. Iscan, F. Rossetto, Optimal time and rate allocation
for a network-coded bidirectional two-hop communication, in
\emph{Proc. IEEE European Wireless Conference (EW)}, Lucca, Italy,
April 2010.


\end{thebibliography}
\end{document}